\documentstyle[epsfig]{l-aa}

\newcommand{\Hal}{\mbox{H$\alpha$}}
\newcommand{\Hbet}{\mbox{H$\beta$}}

\newcommand{\hde}{\mbox{H$_2$}}
\newcommand{\Hi}{\mbox{H{\sc i}}}
\newcommand{\Hii}{\mbox{H{\sc ii}}}

\newcommand{\aprio}{{\em a priori}}

\newcommand{\mic}{\mbox{$\mu$m}}

\newcommand{\mjysec}{\mbox{mJy~arcsec$^{-2}$}}

\newcommand{\dzm}{\mbox{12~$\mu$m}}
\newcommand{\vtm}{\mbox{25~$\mu$m}}

\newcommand{\ctm}{\mbox{100~$\mu$m}}

\newcommand{\fdz}{\mbox{$f$(12~${\mu}$m)}}
\newcommand{\fvt}{\mbox{$f$(25~${\mu}$m)}}
\newcommand{\fsx}{\mbox{$f$(60~${\mu}$m)}}
\newcommand{\fct}{\mbox{$f$(100~${\mu}$m)}}

\newcommand{\hen}{He~2-10}
\newcommand{\Av}{\mbox{$A_{\rm V}$}}

\begin{document}

\thesaurus{03(07.03.1; 09.02.1; 09.12.1; 09.13.1; 19.37.1)}

\title{10~$\mu$m imaging and HI observations of the 
Blue
Compact Dwarf Galaxy He 2-10\thanks{based on data obtained at the
Canada-France-Hawaii Telescope, at E.S.O. La Silla, and at NRAO Green 
Bank}} 

\author{M. Sauvage\inst{1} \and T.X. Thuan\inst{2,1} 
\and P.O. Lagage\inst{1}}

\offprints{msauvage@cea.fr, {\em or visit:} 
http://dphs10.saclay.cea.fr/Publications/publications.html}

\institute{CEA, DSM, DAPNIA, Service d'Astrophysique, C.E. Saclay, 
F91191
Gif-sur-Yvette CEDEX, FRANCE \and Astronomy Department, University of
Virginia, P.O. Box 3818, University Station, Charlottesville, VA
22903-0818, U.S.A.}

\date{received 24/10/96; accepted 18/03/97}

\maketitle
\markboth{M. Sauvage et al.: MIR and \Hi\ observations of the BCD
\hen}{M. Sauvage et al.: MIR and \Hi\ observations of the BCD \hen}

\begin{abstract}

We have observed the Blue Compact Dwarf (BCD) galaxy \hen\ in the 
$\sim$10~\mic\ mid-infrared (MIR) atmospheric window using broad-band 
filters centered at $\lambda10.1$\,\mic\ and $\lambda11.65$\,\mic.  In 
both filters, only the galaxy's central regions were detected.  One of 
the UV emitting regions is not detected, implying an older age.  The 
central region contain two resolved components which have the same MIR 
properties but different \Hal\ fluxes.  We interpret these properties 
in terms of differing star forming histories.  Based on its 
morphology, we show that the MIR emission is unambiguously associated 
with the young massive stars.  We study the spatial variations of the 
MIR color and conclude that they imply the existence of a hot grain 
contribution to the 11.65~\mic\ flux in the central regions of the 
starburst.  We also present a new single-dish measurement of the \Hi\ 
content of He 2--10.


\end{abstract}

\section{Introduction}

Blue compact dwarf (BCDs) galaxies are low-luminosity ($M_{\rm 
B}\ge-18$) systems undergoing intense bursts of star formation in a 
very compact region, as evidenced by their blue UBV colors, and their 
optical spectra which show strong narrow emission lines superposed on 
a stellar continuum rising toward the ultraviolet, similar to spectra 
of \Hii\ regions (Thuan 1991).  Their high \Hi\ content (Thuan \& 
Martin 1981) and their low metallicity 
($Z_{\odot}/50{\la}Z{\la}Z_{\odot}/3$) suggest that BCDs are 
relatively ``young'' systems from the point of view of chemical 
evolution.  Despite their low metallicity and dust content, IRAS has 
shown some BCDs to be strong mid and far infrared emitters (Kunth \& 
Sevre, \cite{ks85}, Sauvage 1991), their dust being heated by an 
intense interstellar radiation field from the many massive stars in 
the starburst regions.

We present here 10~\mic\ imaging and \Hi\ observations of the BCD 
\hen, which Conti (1991) has called a prototypical Wolf-Rayet galaxy.  
This galaxy was chosen because it has one of the highest IRAS \dzm\ 
flux densities among BCDs (Sauvage, \cite{these}).  Moreover optical 
images (Corbin et al., 1993, hereafter CKV) and HST UV maps (Conti \& 
Vacca 1994) have recently been obtained, to which the mid-infrared 
(MIR) observations can be compared.  \Hi\ and CO maps of \hen\ have 
also been recently published by Kobulnicky et al.  (1995).  Our MIR 
observations complement work at other wavelengths in several ways: 1) 
they suffer from little extinction as compared to the visible, 2) they 
have a much higher resolution ($\sim1''$) than the IRAS observations 
(around 1$'$), and 3) they allow to study the spatial distribution of 
the very hot dust as compared to that of the ionizing stars.  We shall 
adopt hereafter a distance of 8.7~Mpc for \hen\ (Tully 1988).  At this 
distance, 1$''$ corresponds to 42~pc.

\section{Observations}

\subsection{MIR imaging}

We have obtained three sets of observations for \hen.  The BCD was 
first observed at the 3.6~m Canada-France-Hawaii Telescope (CFH) on 
April~22, 1992 with the Saclay-built CAMIRAS equipped with a 
64$\times$64 LETI/LIR Si:Ga array (Lagage et al., 1992) through the N 
(8-13~\mic) filter.  With a pixel size of 0$\farcs$45~pixel$^{-1}$, 
CAMIRAS has a field of view of $\sim29''\times29''$.  The effective 
on-source time was 1130~s with an equal time on the OFF position to 
substract the background.  The weather was fair and the Point Spread 
Function (PSF) derived from observations of the calibration star 
$\mu$UMa before and after the galaxy exposure is 1$\farcs$1 Full Width 
at Half Maximum (FWHM).  \hen\ was reobserved in January 1993, during 
an effective on-source time of 1500~s, at the ESO 3.6~m telescope with 
TIMMI, the ESO 10\,\mic\ camera built at Saclay (Lagage et al., 1993), 
through the 11-12.5~\mic\ filter (hereafter referred to as the 
11.65~\mic\ filter).  The scale was 0$\farcs$31~pixel$^{-1}$, giving a 
field of view of $\sim20''\times20''$.  The weather was 
non-photometric and the PSF, determined from reference stars 
observations was 0$\farcs$9 FWHM. We reobserved the galaxy at ESO on 
April 2, 1996 in the same configuration with photometric weather this 
time.  The effective on-source time was 1355\,s and the PSF 1\farcs5 
FWHM.

During that last ESO run, target acquisition (\hen\ and calibration 
stars) was performed by first slewing to a nearby SAO or Bright Star 
Catalog star with accurate coordinates, performing a re-adjustment of 
the telescope control system, and then offsetting to the target. We 
check using the calibration stars that the error in target position is 
always kept below 1$''$.

The observations were done using the standard chopping and nodding 
techniques.  Typically 20 frames of 7.7\,ms integration time each are 
taken in one position of the chopping mirror and summed.  We perform 
20 chopping cycles and subtract the OFF from the ON positions to 
create an image which is stored.  Ten such images are created before 
the telescope is moved so that the ON and OFF beams are inverted.  The 
number of such nodding cycles as well as the individual integration 
time are modified according to the intrinsic flux of the source and 
the desired signal-to-noise ratio.  The chopping motions allow to 
substract the thermal emission from the sky as well as the spatially 
homogenous part of the telescope emission, while the nodding motions 
allow to remove residual emission from the telescope structure due to 
slight differences in the optical path for the two positions of the 
chopping mirror, as well as slow sky variations.  The data were 
reduced using a software based on IDL developed at Saclay.  Prior to 
any noise filtering processing, peak signal-to-noise ratios are 5.4 
for the CFHT image, 7.2 for the 01/93 ESO image, and 6.5 for the 04/96 
ESO image.  The better S/N obtained during the 01/93 run is due to 
both the longer integration time and the use of a smaller pixel field 
of view that allows to operate the detector in nominal conditions.

The N-band filter has a central wavelength of 10.1~\mic\ with 
half-power wavelengths of 7.7 and 12.8\,\mic\ (the transmission curves 
of all filters used here can be found on the ESO server at {\tt 
http://http.hq.eso.org/proposals/ timmi.html}, the CFHT system being 
identical to the ESO one).  Our image was calibrated using the 
standard star $\mu$UMa whose magnitude at 10.1~\mic\ is -1.03 
(Tokunaga 1984).  Since the 10.1\,\mic\ flux density of a 0~mag star 
is 38~Jy (Abbot et al., 1984), assuming a spectral energy distribution 
going as $f_{\nu} \propto \nu$ (see section 5.2), the total N-band 
flux density of \hen\ is 1060~mJy, with an uncertainty of $\sim$10\%.  

The 11.65 filter has a central wavelength of 11.65\,\mic\ with 
half-power wavelengths of 10.8 and 12.5\,\mic.  To calibrate the 
11.65\,\mic\ image we used the stars $\lambda$Vel (N$=-1.78$) and 
$\alpha$Cen\,a (N$=-1.56$).  We computed the expected flux densities 
of these calibrators at 11.65\,\mic\ assuming a stellar spectral 
energy distribution.  Assuming again a spectral energy distribution 
going as $f_{\nu} \propto \nu$, we find a total 11.65\,\mic\ flux 
density of 850\,mJy, with an uncertainty of 15\%.

Two previous MIR photometric measurements of \hen\ have been 
published.  A low resolution map has been obtained by Telesco et al.
(1993) who measured a 10.8\,\mic flux density of only 600~mJy, with an 
uncertainty of 10-15\%.  As Telesco et al. (\cite{telesco93}) made 
no spectral energy distribution correction for their measurement, i.e.  
they assumed a stellar energy distribution ($f_{\nu} \propto \nu^{2}$) 
for both the calibrators and the galaxy, we need to derive ours in the 
same conditions prior to any comparison.  With a stellar energy 
distribution, our 10.1\,\mic\ flux density is only 970~mJy.  This 
measure predicts a flux density of 850~mJy at 10.8\,\mic.  This is 
only marginally consistent with the measurement of Telesco et al.  
(\cite{telesco93}): 850~mJy~$-$~15\% is 720~mJy while 600~mJy~$+$15\% 
is 690~mJy.  The remaining discrepancy can probably be explained by 
the fact that (1) their detector pixels are spaced by more than their 
FWHM and thus do not fully sample the sky, and (2) that \hen\ is 
nearly resolved in their observation so that the correction to apply 
to the detected flux to obtain the true flux is difficult to deduce 
from the calibration star observations.  If we assume that the signal 
loss is proportional to the sky undersampling, then the flux densities 
are in agreement.  Cohen \& Barlow (1974) have also observed \hen\ in 
the N-band.  They report a N magnitude of 3.5$\pm$0.2 within a 11$''$ 
diameter beam, which translates into a flux density of 1.6~Jy, higher 
than either our or Telesco et al.  measurements.  We believe that the 
difference cannot be due to a low level diffuse component as the more 
sensitive \dzm\ IRAS measurement, with a beam covering the entire 
galaxy rather than just the star-forming regions, gives only a flux 
density of 1.1~Jy, lower than the Cohen \& Barlow measurement, even 
after correction for the different bandwidths and taking into account 
the assumed IRAS spectral shape (see also our discussion in 
\ref{sec:dust}).  Furthermore, using our images, we have simulated 
aperture photometry with an 11$''$ aperture and find no significant 
low-surface brightness component.  We thus believe that measurement to 
be in error.

\begin{figure*}
\hbox{
\epsfxsize=9cm
\epsfbox{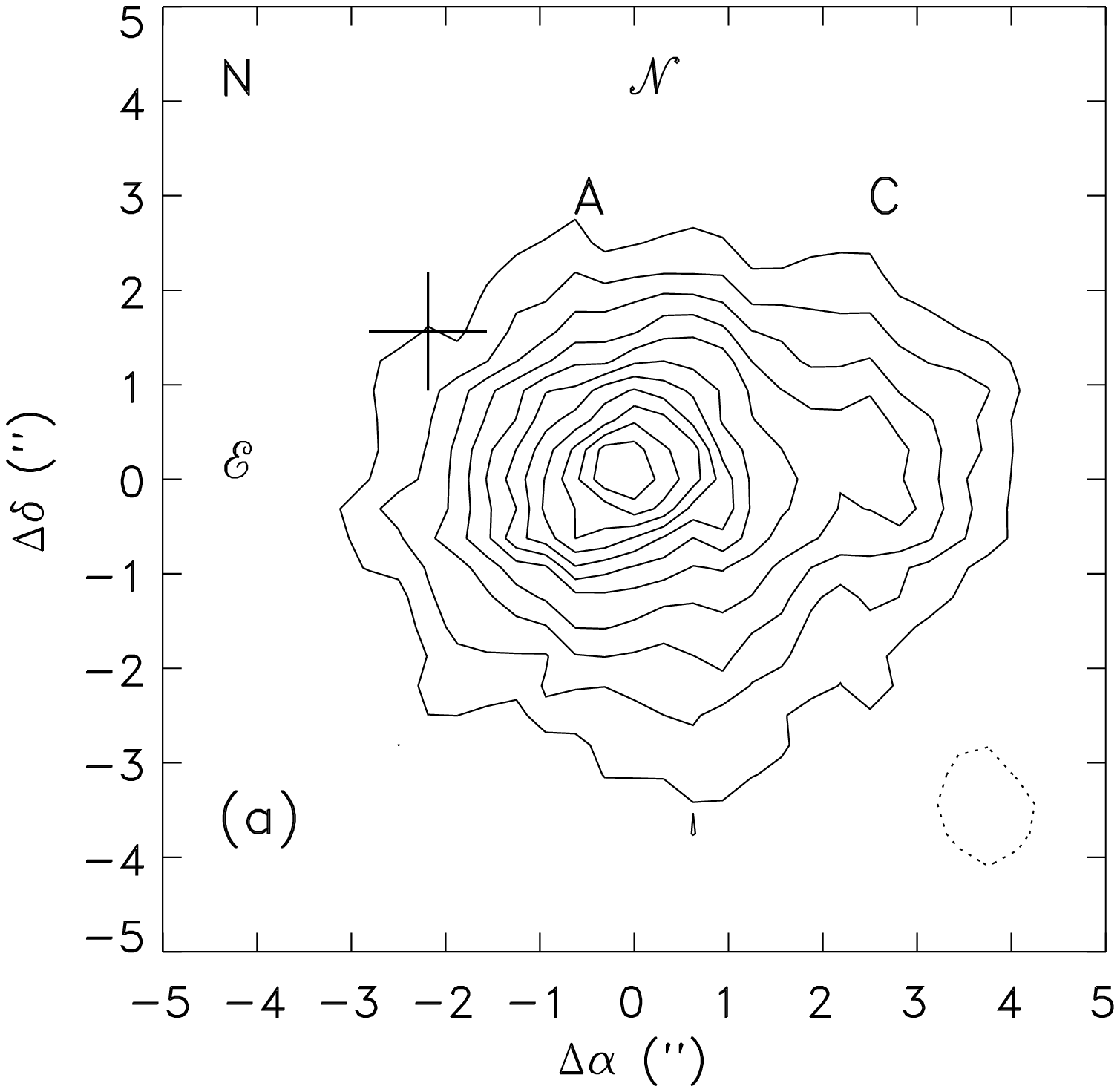}
\epsfxsize=9cm
\epsfbox{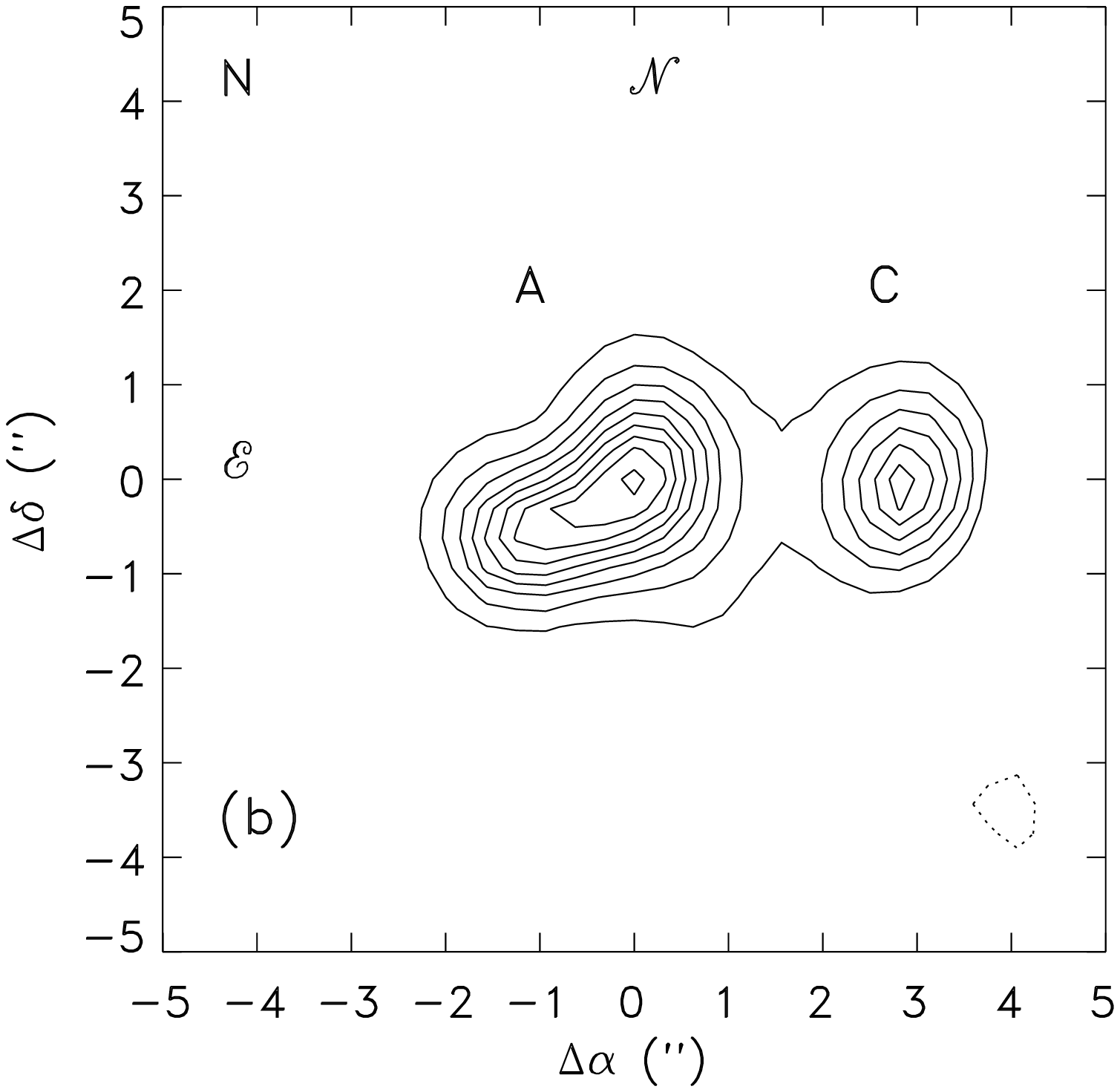}
}

\caption[ ]{\hen\ in the N-band filter, with labels of the 
star-forming regions following the nomenclature in the text.  (a) 
Original image filtered with the multiscale, maximum entropy method of 
Pantin \& Starck (\cite{pantin96}).  Contours start at 6$\sigma$ and 
are spaced by 2$\sigma$ (1$\sigma$ is 2.2~\mjysec).  The dotted 
contour at bottom right is the FWHM contour (1$\farcs$1) of the PSF 
after a similar treatment.  A cross in the upper left corner indicate 
the location of the CO peak observed by Kobulnicky et al.  
(\cite{kobul}).  (b) Deconvolved image.  The FWHM is now $0\farcs8$, 
shown as a dotted contour in the bottom right corner, and contours 
start at 5$\sigma$ (as measured in the deconvolved image) and are 
spaced by 3$\sigma$.}

\label{CFH}
\end{figure*}

\begin{figure*}
\hbox{
\epsfxsize=9cm
\epsfbox{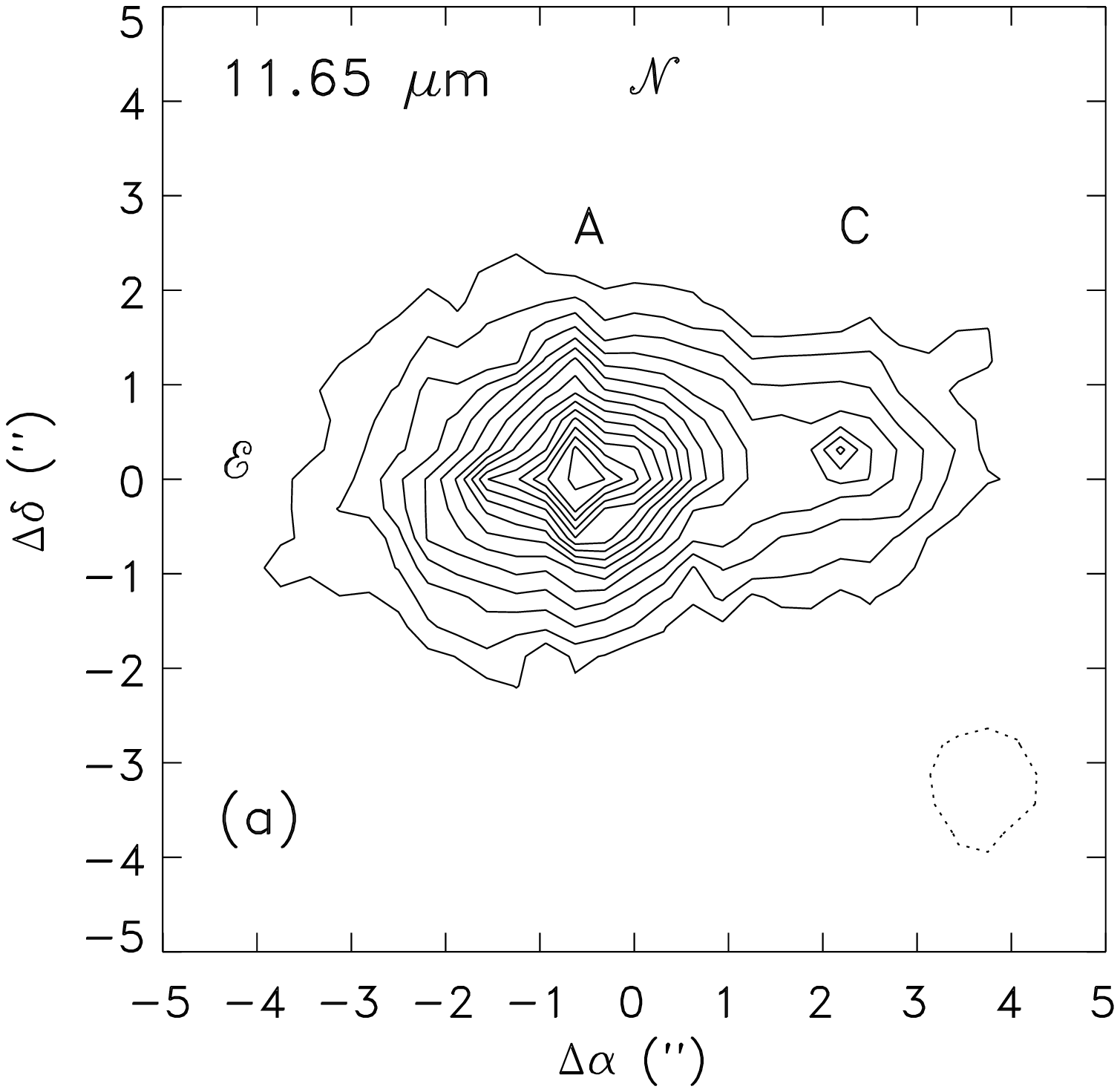}
\epsfxsize=9cm
\epsfbox{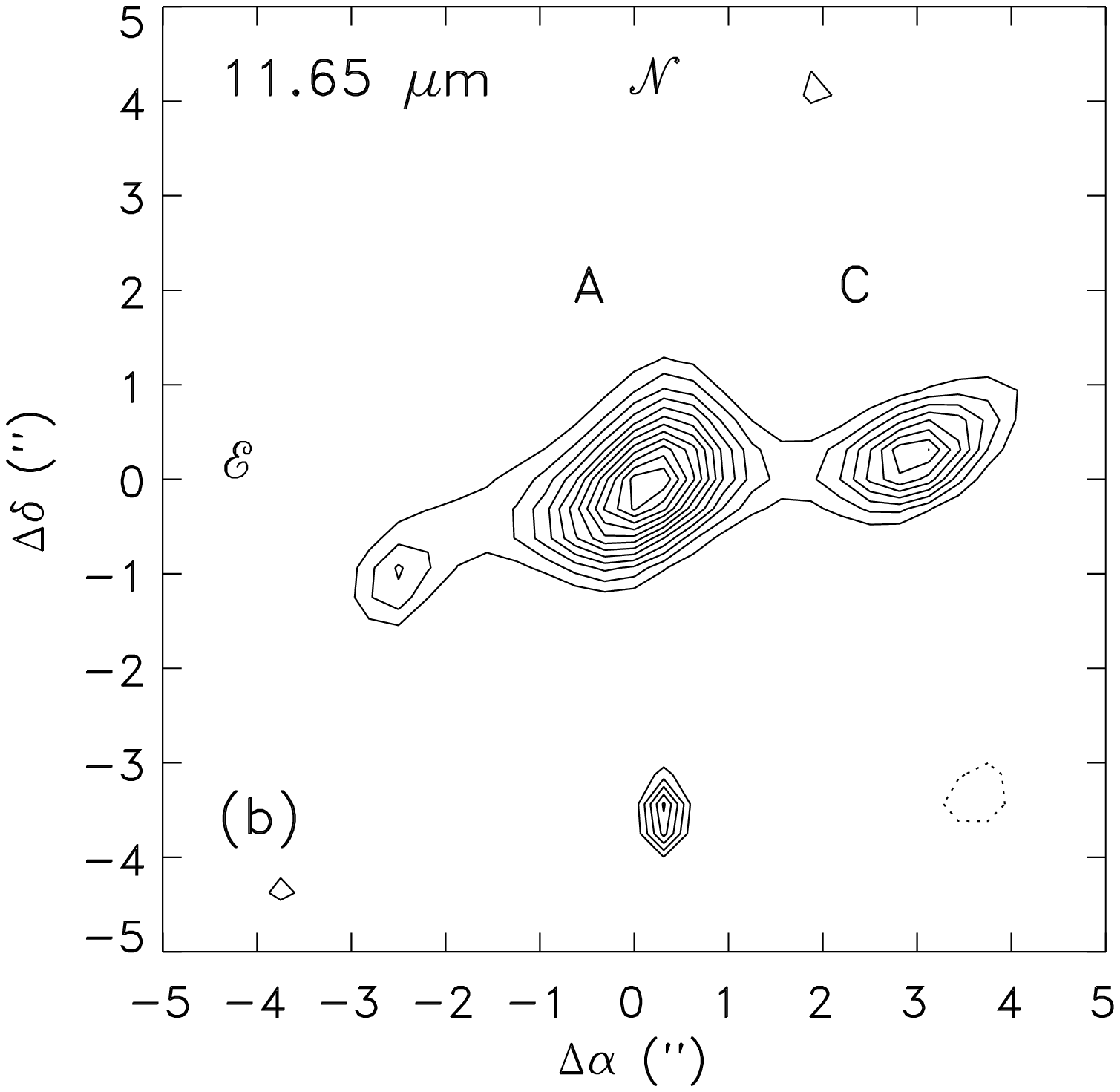}
}

\caption[ ]{\hen\ in the 11.65~\mic\ band filter, with labels of the 
star-forming regions following the nomenclature in the text.  (a) 
Original image filtered with the multiscale, maximum entropy method of 
Pantin \& Starck (\cite{pantin96}).  Contours start at 12$\sigma$ and 
are spaced by 6$\sigma$ (1$\sigma$ is 1~\mjysec).  The dotted contour 
at bottom right is the FWHM (0$\farcs$9) of the PSF after the same 
filtering treatment.  (b) Deconvolved image.  The FWHM is now 
$0\farcs5$, shown as a dotted contour in the bottom right corner, and 
contours start at 5$\sigma$ (as measured in the deconvolved image) and 
are spaced by 3$\sigma$.}

\label{ESO}
\end{figure*}

Figures \ref{CFH}a and \ref{ESO}a show respectively the N and 
11.65~\mic\ images filtered with the multiscale maximum entropy method 
of Pantin \& Starck (\cite{pantin96}).  The peak signal to noise ratio 
is now 30 in figure~\ref{CFH}a and $\sim$100 in figure~\ref{ESO}a.  
The initial difference in signal-to-noise ratio is amplified by the 
filtering process and also probably by the more compact nature of the 
emission at 11.65~\mic.  Note that the first contour in both images 
corresponds to approximately the same brightness level.

Spatial information at sub-arcsecond scale can still be restored with 
a deconvolution algorithm based on the same techniques.  Figures 
\ref{CFH}b and \ref{ESO}b show the deconvolved images.  The FWHMs of 
these images are respectively 0$\farcs$8 and 0$\farcs$5 for the N and 
11.65~\mic~images.  Although they appear at a relatively high $\sigma$ 
level, the structures seen East and South of region~A in 
figure~\ref{ESO}b are very likely artefacts of the deconvolution 
procedure.

\subsection{21~cm observations and the gaseous component}

\begin{figure}[ht]
\epsfxsize=9cm
\epsfbox{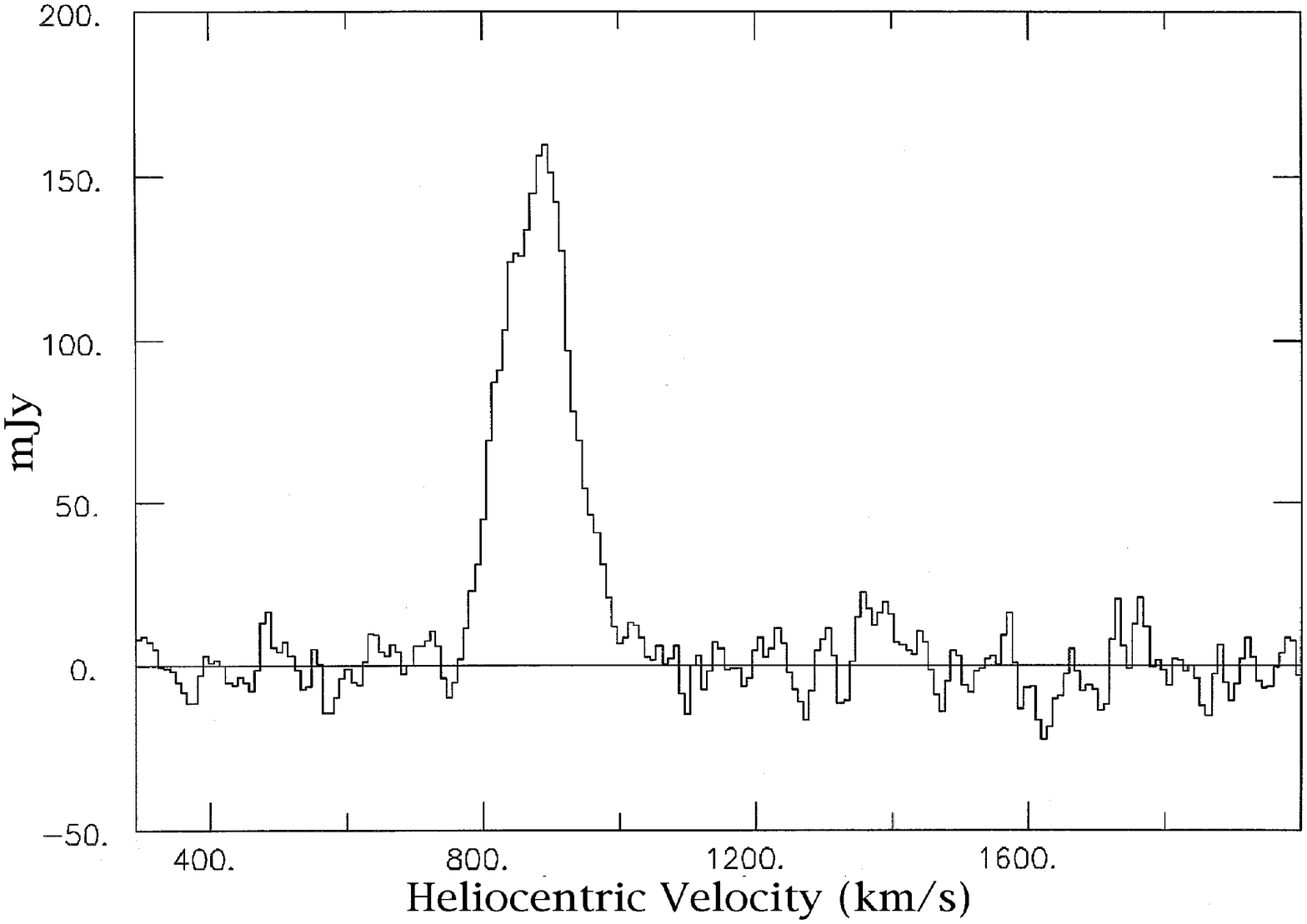}

\caption[ ]{21 cm profile of \hen\ obtained with the NRAO 43~m 
telescope at Green Bank.}

\label{HI}
\end{figure}

\Hi\ observations of \hen\ were made with the NRAO\footnote{The 
National Radio Astronomy Observatory is operated by Associated 
Universities, Inc., under contract with the National Science 
Foundation.} 43~m telescope at Green Bank on Nov~26, 1994, with a 
two-channel, dual polarization 21~cm prime-focus receiver with a 
system temperature of $\sim$20~K. The half-power beam width of the 
telescope is $\sim22'$. A bandwidth of 20~MHz was used with 
a 1024 channel autocorrelator split in two.  The two polarizations 
were detected independently and averaged to improve sensitivity.  The 
observations were made in the total-power position-switching mode, 
with 6~mn on-off integrations.  Taking into account the two 
polarizations, a total integration time of 72~mn on source was 
obtained.  The resulting spectrum is shown in figure~\ref{HI}.  The 
effective resolution after Hanning smoothing is 16~km~s$^{-1}$ and a 
gain of 3.4~Jy/K was adopted.  The derived \Hi\ parameters are: the 
heliocentric velocity $v_{\rm H}=883\pm5$~km~s$^{-1}$, the velocity 
width at 20\% of peak intensity ${\Delta}v_{20}=181\pm16$~km~s$^{-1}$, 
the velocity width at 50\% of peak intensity 
${\Delta}v_{50}=119\pm10$~km~s$^{-1}$ and the neutral hydrogen flux 
$F_{\rm \Hi}=17.09\pm1.04$~Jy~km~s$^{-1}$.  With a distance of 
8.7~Mpc, the \Hi\ mass is 3.05$\pm0.19~10^{8}$~M$_{\odot}$.  The peak 
flux density is 163~mJy, the rms noise is 8.8~mJy and the errors in the 
velocity, velocity width, and flux density are calculated as in 
Schneider et al. (1990).

\hen\ has been previously observed in \Hi\ by Allen et al. (1976) 
using the Parkes 64~m telescope with a 15$'$ half-power beam width 
(HPBW).  While there is a good agreement for $v_{\rm H}$ and the \Hi\ 
flux (the Allen et al. values are respectively 880$\pm$20 km 
s$^{-1}$ and 17.26 Jy km s$^{-1}$), our ${\Delta}v_{50}$ is smaller 
and our peak flux density is higher (Allen et al.  found 160$\pm$20 km 
s$^{-1}$ and 110~mJy).

Kobulnicky et al.  (1995) have recently obtained a VLA \Hi\ map of He 
2--10.  They obtained an integrated \Hi\ flux of 9.87$\pm$0.13 Jy km 
s$^{-1}$, only 58\% of our value and a peak flux density of 90 mJy, only 55\% 
of our value.  As Kobulnicky et al., we interpret these differences as 
due to a low-intensity smooth outlying \Hi\ component resolved out by 
the interferometer.  An extended population of 10$^5$ to 10$^6$ 
M$_\odot$ \Hi\ clouds lying in the periphery of the BCD will be 
missed.  The VLA map gives a line center of 873$\pm$10 km s$^{-1}$ and 
a ${\Delta}v_{50}$ of 106$\pm$10 km s$^{-1}$, in good agreement with 
our values.  The \Hi\ velocity field as seen with the VLA is 
consistent with solid body rotation, which results in a nearly 
gaussian shape of the \Hi\ single-dish profile as seen in figure 3.  
However, there is a definite asymmetry in the \Hi\ profile.  There is 
a low-intensity wing on the high-velocity side which is not present on 
the low-velocity side.  This asymmetry is also present in the VLA 
synthetic single-dish spectrum of Kobulnicky et al.  (1995).  
Examination of the VLA \Hi\ total intensity and velocity maps shows 
that the high velocity component may be due to a distinct \Hi\ cloud 
to the north of the BCD. Solid body rotation in the interior part 
implies a mass distribution with a constant density in the inner 1.5 
kpc radius.  Adopting $B_{\rm T}$~=~12.45 (de Vaucouleurs et al., 
1991), we obtain the distance-independent hydrogen mass to blue 
luminosity ratio $M(${\Hi}$)/L_{\rm B}$~=~0.25\, in solar units (to 
allow a direct comparison of this ratio to that measured for other 
galaxies, we have adopted here the convention expressing $L_{\rm B}$ 
in units of the Sun's blue luminosity, $L_{\rm B}^{\odot}$; note that 
this differs from the convention used in Table~1, and that the ratio 
$L_{\odot}/L^{\odot}_{\rm B}$ is $\sim6.25$).

Kobulnicky et al.  (1995) also obtained a $^{12}$CO (1--0) map of the 
BCD. While the \Hi\ envelope is much more extended than the 
low-surface-brightness underlying elliptical stellar component (with 
colors typical of K-giant stars, Corbin et al., \cite{corbin}) of 
the BCD, the CO emission is confined to a region of size comparable to 
that of the emission-line regions (see Corbin et al, \cite{corbin}, 
their figure 2c), with however quite a different shape.  The peak CO 
and \Hi\ brightnesses are separated by $\sim$500 pc.  The Wolf-Rayet 
starburst is located between these two peaks, closer to the CO peak 
(see figure~\ref{CFH}a).  The \Hi\ map shows a hydrogen deficiency in 
the vicinity of the two starburst regions, which Kobulnicky et al.  
suggest may be the result of a replacement effect by molecular gas 
which fills this region.  The CO data imply a total molecular gas mass 
of 1.6$\pm$0.1 $\times$ 10$^8$ M$_\odot$, assuming a Galactic 
CO-to-H$_2$ conversion factor.  Thus the total gaseous mass is 4.7 
$\times$ 10$^8$ M$_\odot$.  The CO rotation curve implies that the 
mass of the young UV bright ``proto globular clusters" discussed by 
Conti \& Vacca (1994) can comprise up to half of the dynamical mass in 
the inner 70 pc radius.  Both \Hi\ and CO maps show a tidal tail in 
the SE direction, suggesting that He 2-10 is the product of a merger 
between two dwarf galaxies.  However, the smooth r$^{1/4}$ profile of 
the underlying low-surface-brightness extended component (CKV) implies 
that it is a moderately advanced merger as the underlying stellar 
component has had time to relax.  The total dynamical mass implied 
from the \Hi\ velocity field is 2.7 $\times$ 10$^9$ M$_\odot$/ 
$(sin\,i)^2$ where $i$ is the inclination angle of the BCD. Baas et 
al.  (1994) have obtained a dynamical mass of 6.7$\pm$3.4 $\times$ 
10$^9$ M$_\odot$ by extrapolation of the K-band flux.  This would 
correspond to $i$ = 39$^{\circ}$.  This inclination angle would give a 
M$_{dyn}$/L$_B$ ratio of $\sim$ 5.  Thus, unless the inclination is 
very low ($i$ $\leq$ 10 degrees), M$_{dyn}$/L$_B$ is relatively small, 
implying that most of the matter in He 2-10 is not dark but luminous 
(Lo et al., \cite{loSY}).  The corresponding gaseous mass fraction 
would be 0.07, somewhat on the low side of values for BCDs (Thuan 
1985), a conclusion also reached by Baas et al.  (1994).

\section{MIR morphology}

\begin{table*}
\caption[ ]{Observed and derived data on \hen.}
\begin{minipage}[t]{18cm}
\begin{flushleft}
\begin{tabular}{lrlr}
\hline
\multicolumn{2}{c}{Observed quantities} & \multicolumn{2}{c}{Derived 
quantities} \\
\hline
$\alpha_{1950.0}$ 	& $08^{\rm h}34^{\rm m}07\fs1$ 	& 			& 			\\
$\delta_{1950.0}$ 	& $-26^{\circ}14'04''$ 		& 			& 			\\
redshift (km~s$^{-1}$) 	& 883$\pm$5 			& Distance (Mpc) 	& 8.7 			\\
$D_{25}$ ($''$) 	& 104$\pm$10 			& 			& 			\\
$R_{25}$ ($''$) 	& 70 				& 			& 			\\
$m_{\rm B}$ 		& 12.45 	& $L_{\rm B}$
					\footnote{Here $L_{\rm B}$ is 
					expressed in bolometric solar 
					units, not in units of the Sun's 
					blue luminosity. The ratio 
					$L_{\odot}/L_{\rm B}^{\odot}$ is $\sim6.25$~.} 
					(L$_{\odot}$) 		& 1.91~$10^{8}$ 	\\
\fdz\ (Jy) 		& 1.1 		& $L_{12~{\mu}{\rm m}}$  (L$_{\odot}$) 		& 
3.5~$10^{8}$ \\
\fvt\ (Jy) 		& 6.55 		&  $L_{25~{\mu}{\rm m}}$  (L$_{\odot}$) 	& 
8.1~$10^{8}$ \\
\fsx\ (Jy) 		& 23.8 		&  $L_{60~{\mu}{\rm m}}$  (L$_{\odot}$) 	& 
1.47~$10^{9}$ \\
\fct\ (Jy) 		& 25.7 		&  $L_{100~{\mu}{\rm m}}$  (L$_{\odot}$) 	& 
6.14~$10^{8}$ \\
 			& 		& $L_{\rm FIR}$ (L$_{\odot}$) 			& 2.6~$10^{9}$ \\
$f({\rm N})$ (Jy) 	& 1.06$\pm$0.1 			& $L_{\rm N}$ (L$_{\odot}$) 	& 
(3.6$\pm$0.4)~$10^{8}$ \\
$f^{\rm A}({\rm N})$ (Jy) 	& 0.71 			& 				& \\
$f^{\rm B}({\rm N})$ (Jy) 	& $<0.1$ 		& 				& \\
$f^{\rm C}({\rm N})$ (Jy) 	& 0.35 			& 				& \\
$f(11.65~\mu{\rm m})$ (Jy) 	& 0.85$\pm$0.1 		& $L_{\rm 11.65}$ 
(L$_{\odot}$) & (1.2$\pm$0.2)~$10^{8}$ \\
$f^{\rm A}(11.65~\mu{\rm m})$ (Jy) 	& 0.57 		& 				& \\
$f^{\rm B}(11.65~\mu{\rm m})$ (Jy) 	& $<0.1$ 	& 				& \\
$f^{\rm C}(11.65~\mu{\rm m})$ (Jy) 	& 0.28 		& 				& \\
$F_{\rm H{\sc i}}$ (Jy~km~s$^{-1}$)\footnote{This paper.}
                            		& 17.09$\pm$1.04 	& $M_{\rm H{\sc i}}$ 
(M$_{\odot}$) 		& (3.1$\pm$0.19)\,10$^{8}$ \\
$\Delta_{v_{50}}$ (km~s$^{-1}$)$^{b}$
                                	& 119$\pm$10 	&						& \\
			& 		& $M_{\rm H_{2}}$ (M$_{\odot}$)\footnote{Kobulnicky et al. 
                    (\cite{kobul}), adopting a Galactic CO-to-\hde\ 
                    conversion factor.}
                                    & 1.6$\pm$~$10^{8}$ \\
            &       & $M_{\rm gas}$ (M$_{\odot}$)           & 
4.7~$10^{8}$ \\
			& 		& $M_{\rm total}$ (M$_{\odot}$)\footnote{Kobulnicky et al. 
			(\cite{kobul}), where $i$ is the inclination angle. Baas et al. 
			(\cite{baas}) found 6.7$\pm$3.4\,10$^{9}$\,M$_{\odot}$, from the 
			extrapolated K-band flux, 
			corresponding to $i = 39\deg$.} 		& 2.7~$10^{9}$/($\sin i)^{2}$ \\
			& 		& $M_{\rm dust}(IRAS)$ (M$_{\odot}$) 		& 7.6~$10^{4}$ \\
			& 		& $M_{\rm dust}(\tau_{\rm Si})$ (M$_{\odot}$) 	& 
$<10.3~10^{4}$ \\
			& 		& $M_{\rm H{\sc i}}/L_{\rm B}$
					\footnote{In units of the Sun's 
					blue luminosity.} 				& 0.25 \\
			& 		& $M_{\rm dust}/M_{\rm gas}$ 		& $\la2.2~10^{-4}$ \\
			& 		& $M_{\rm gas}/M_{\rm total}$ 		& 0.17 ($\sin i)^{2}$ \\
\hline
\end{tabular}
\end{flushleft}
\end{minipage}
\end{table*}

\hen\ has been extensively imaged in the visible, both in broad and 
narrow band filters (Hutsemekers \& Surdej 1984, CKV).  The $B$ and 
$V$ images show high surface brightness irregular~(i) star-forming 
regions in the central part of a more extended 
lower-surface-brightness elliptical~(E) envelope, typical of the iE 
morphology defined by Loose \& Thuan (1985).  The color of the 
elliptical envelope is consistent with that of K giant stars (CKV).

Three starburst regions can be distinguished on the optical images.  
Following the nomenclature of CKV, the most prominent is region~A at 
the center of the galaxy.  Elongated in the SE-NW direction by 
$\sim6''$ (FWHM) on $B$ and $V$ images, it is pointlike with a diffuse 
halo on \Hal\ images (CKV).  The second region is B, located at 
$\sim8\farcs5$~E of A. It is smaller (5$''\times2''$) and fainter in $B$ 
and $V$ images and is seen at an even lower level in \Hal.  The third 
region, called C here, located 2$\farcs$7~W of region A is not 
prominent in $B$ and $V$ images but is more noticeable in \Hal\ and still 
more noticeable in [O{\sc iii}].

In the MIR, only regions A and C can be seen.  Region~B, although 
within the field of view of the cameras, was not detected.  We 
establish 3$\sigma$ upper limits at the optical location of B of 
8~\mjysec\ in the N-band and of 3~\mjysec\ in the 11.65~\mic\ band, 
corresponding to upper limits for the total flux density of 80~mJy in the 
N~band and 30~mJy in the 11.65~\mic~band, if we adopt for region~B a 
MIR extent equal to its optical size of $5''\times2''$.  We note that 
even when these upper limits are taken into account, our N-band 
measurement is still smaller than that of Cohen \& Barlow (1974).  
Furthermore, it is not even clear if region~B was inside the beam used 
by these authors.

The MIR morphologies are very similar in both N and 11.65~\mic\ 
images: regions A and C are clearly resolved and separated.  The 
peak-to-peak distance is 3$''$, which corresponds to a linear scale of 
126~pc.  For region~A the position angle remains approximately the 
same from N to 11.65~\mic.  This is not the case for region~C, where 
there is a 45-90$^{\circ}$ rotation.  The origin of this difference is 
unclear.

Apart from the extension of A, the MIR images are very similar to the 
\Hal\ image, and even more to the [O{\sc iii}] image of CKV, implying 
that the dust responsible for the MIR emission is either coextensive 
with or closely connected woth the hot ionized gas 
responsible for the \Hal\ and [O{\sc iii}] emission.  Our MIR images 
do not show evidence for a dust lane crossing the BCD North-South and 
separating regions A and B as hypothesized by Allen et al.  (1976) and 
Hutsemekers \& Surdej (1984).

Conti \& Vacca (1994) have recently obtained 2200~\AA\ images of \hen\ 
with the Hubble Space Telescope (HST), where the emission regions are 
clearly resolved into sub-components.  Region~A is resolved into 
$\sim$10 bright knots forming an arc, interpreted by the authors as 
proto-globular clusters (age$\la$10~Myr).  The SE-NW orientation of 
the arc is the same as that of region~A in the MIR. The arc is 
$\sim2''$ long in the EW direction, thus is completely within 
region~A. Region~B is also resolved into several subcondensations, but 
instead of being ordered into an arc, the knots are scattered randomly 
throughout the region.  The total UV luminosity of the knots in 
region~B is 4 times fainter than that in region~A. Region~C is 
resolved into 3 knots with a total $L_{\lambda}$(2200~\AA) luminosity 
about 20 times fainter than that of region~A. It is interesting to 
note that the structure of region~A observed in the UV by Conti \& 
Vacca (1994) is more elongated than the \Hal\ structure seen by CKV. 
The latter authors derive a diameter of $\la$34~pc for the ionizing 
source, while the arc of stellar clusters observed by Conti \& Vacca 
is $\sim$85~pc long.  A few clusters are resolved in the UV and their 
diameter is $\sim$10~pc.  It thus appears that only 1 or 2 of the UV 
clusters still contain massive ionizing stars.

Using the interferometric CO map of Kobulnicky et al.  (\cite{kobul}), 
we can obtain a better picture of the starburst in \hen\ by comparing 
the spatial distributions of the CO and MIR emissions.  In 
figure~\ref{CFH}a we show the position of the CO peak.  It is located 
NE from the MIR peak, just on the lowest contour.  The distribution of 
the CO gas is also slightly elongated, with the same position angle as 
region~A in the MIR. This suggest that region~A is actually located 
on the SE edge of the densest part of the CO cloud instead of just 
being seen projected against it.  Region~C, further to the W, is in 
a region of lower CO column density, while region B is completely 
outside the CO cloud on the other side.

A plausible way to bring together all the observations of \hen\ in a 
coherent picture is the following: we are seeing in the BCD a 
photodissociation region resulting from the effect of a compact 
star-forming region on a neighboring CO cloud.  Only the one or two 
youngest ($\leq$ 10$^7$ yr) star clusters in the star-forming region 
still contain ionizing stars, which would explain the very compact 
aspect of the galaxy in \Hal.  However all star clusters are still 
young enough ($\leq$ 10$^8$ yr) to produce copious amounts of UV 
radiation.  This radiation photodissociates molecules in the nearby CO 
cloud as well as breaking up the larger grains or removing small 
dust grains from the mantles of the larger ones (e.g.  Boulanger et 
al., \cite{boul90}).  \hde\ emission is detected only to the East of 
region A (Baas et al., \cite{baas}) where most of the molecular gas 
is.  \hen\ has also been detected in [C{\sc ii}] with a flux of 
$\sim$2\,10$^{-14}$\,W\,m$^{-2}$ (S. Madden, private communication).  
If the [C{\sc ii}] and the \hde~S(1)\,1-0 emission regions are 
coextensive, then their observed line intensity ratio 
($\sim$10$^{-3}$) can be explained by UV excitation of a gas with 
density $10^{2-3}$\,cm$^{-3}$ and an exciting radiation field 10$^{4}$ 
times more intense than the solar neighborhood radiation field 
(Hollenbach et al., \cite{hollen91}, Burton et al., \cite{burton92}).  
However, given the very different beams used to observe these two 
emissions, it is quite likely that the \hde\ emission is more 
concentrated than the [CII].  In that case shock excitation is 
required to produce the amount of detected \hde\ emission.  Such 
shocks are in fact to be expected in a starburst region.  We note 
however that if the [C{\sc ii}] emission is spread out over the 55$''$ 
beam used to observe \hen, then its brightness is that expected from a 
classical photodissociation region with the density and radiation 
conditions mentioned above.

Thus this picture of a photodissociation region created by a very 
compact starburst located at the edge of a dense molecular cloud 
provides a reasonable description of all data available to date on the 
starburst in \hen.  In that picture, the elongated morphology of 
region~A in the MIR can be understood if the emitting dust is mixed 
with the UV clusters seen by Conti \& Vacca (\cite{cova}) or if the 
emission actually from the interface between the ionized region and 
the large molecular complex mapped by Kobulnicky et al.  
(\cite{kobul}), with a geometry reminiscent of the Orion Bar (e.g.  
Tielens et al., \cite{tielens93}, or Bregman et al., 
\cite{bregman89}).  This geometry is in fact favored by the 
observation that there does not seem to be a large amount of material 
on the line of sight to the UV clusters (see 
section~\ref{sec:extinct}).

\section{Extinction}
\label{sec:extinct}

There exists several extinction estimates for region~A. Using the 
\Hal/\Hbet\ ratio and case~B recombination theory, Sugai \& Taniguchi 
(1992, hereafter ST) derive a total (Galactic $+$ internal) extinction 
of \Av$\sim$0.9, while Vacca \& Conti (1992) obtain \Av$\sim$1.7~mag.  
With the Br$\gamma$ measurement from Moorwood \& Oliva (1988), we 
derive from the Br$\gamma$/\Hbet\ ratio and case B recombination 
theory \Av$\sim$1.2~mag, in good agreement with the optical estimates.  
The extinction from the Milky Way is \Av$\sim0.6$ (Johansson, 1987), 
so that the intrinsic extinction in the optical is $\sim0.6$~mag.

Aitken \& Roche (1984) have, however, derived a much larger extinction 
\Av$\sim10\pm$5~mag, using the optical depth $\tau_{\rm Si}$ of the 
silicate absorption band at 9.7~\mic.  Adopting a Galactic ratio of 
molecular hydrogen column density to visual extinction, Kobulnicky et 
al.  (1995) have derived from their CO map an even higher extinction 
\Av\ $\sim$ 30 mag.  Given that the CO cloud is not coexistent with 
the star-forming region, it is not surprising that the \Av\ derived 
from the CO data does not agree with the {\Av}s derived from optical 
or MIR measurements.  However, the large discrepancy between the 
optical and MIR {\Av}s needs to be discussed.

We first note that the determination of the silicate absorption depth 
is not an easy task.  The existence of two broad emission features on 
both sides of the silicate band (the 7.7 and 8.6\,\mic\ ``blend'' and 
the 11.3\,\mic\ band, see e.g.  the ISOCAM spectrum in Boulade et al., 
\cite{boulade96}) makes it very difficult to locate the true level of 
the continuum, especially on the short wavelength side. The difficulty 
in measuring $\tau_{\rm Si}$ could explain why, in Phillips et al. 
(\cite{phillips84}), five out of seven objects have an \Av\ as 
determined from silicate absorption larger than their \Av\ as derived 
from the Balmer decrement.
  
Alternatively, one could consider these discrepancies as real.  Puxley 
\& Brand (1994) have recently addressed the problem of extinction in 
the starburst galaxies NGC 1614 and NGC 7714 which, like \hen, show 
large discrepancies in extinctions measured at different wavelengths.  
They show that these measures can be reconciled if, instead of 
adopting a uniform foreground obscuring screen for the dust 
distribution, one assumes that the dust distribution is patchy.  In 
that case, the shape of the absorption curve is such as to create an 
increasing bias for low extinction zones as wavelengths become 
shorter.  Indeed, Fanelli et al.  (1988) have found that in BCDs, the 
UV extinction is smaller than that derived from the Balmer decrement.  
They interpret the low UV extinction to be the result of a selection 
effect: UV radiation can only escape from regions where dust 
extinction is not large.  A plausible origin for the patchy dust 
distribution is that the dust is efficiently destroyed or removed from 
the vicinity of O~stars by the combined action of ionizing radiation, 
stellar winds and supernovae.  This idea is supported by the HST 
observations of Conti \& Vacca (1994) who estimate that region~A 
contains $\sim$30000~O~stars within a diameter of $\sim$90~pc 
(Schaerer, \cite{schaerer}, recently showed that this number may in 
fact be a lower limit to the true number of O stars in \hen).  This 
corresponds to a very large energy density, comparable to that in the 
vicinity ($\sim$1.2~pc) of the central ionizing cluster of the Rosette 
nebula in the Milky Way.  For the latter, a drop in the \dzm\ emission 
is observed at that distance, which is presumably due to dust 
destruction by the intense radiation field (Cox et al., 1990).

Therefore, one can assume that in \hen, UV and optical light comes 
directly from the stellar clusters through holes in 
the gas and dust cloud.  The MIR light, emitted preferentially at the 
photodissociation front (Tielens et al., \cite{tielens93}), i.e.  
deeper inside the cloud and projected at a different location on the 
sky, suffers from some extinction by foreground colder dust.

Because of the uncertainty in the true extinction and the poorly 
known shape of the MIR extinction curve, we have chosen not to correct 
the MIR measurements discussed here for extinction.

We note however that if we use $\tau_{\rm Si}$= 0.7$\pm$0.3 (Aitken \& 
Roche 1984) and the silicate absorption coefficient of Draine \& Lee 
(1984), we derive a silicate mass of 5.1$\pm2.2\,10^{4}$\,M$_{\odot}$.  
If we further assume that silicates make up approximately one half of 
the total dust mass, we obtain a mass of $\sim$ 10.2 
$10^{4}$\,M$_{\odot}$, comparable to the value of $\sim$ 
7.6\,$10^{4}$\,M$_{\odot}$ estimated from the IRAS FIR flux densities.  This 
would correspond to a dust-to-gas mass (\Hi+H$_2$) ratio of $\sim$ 2.2 10$^{-4}$, 
as compared to 7 10$^{-3}$ in the solar neighborhood.  Such a low 
value for the dust-to-gas mass ratio is not unusual for late-type 
dwarfs (Sauvage \& Thuan 1994).

\section{The interstellar medium and star-forming properties of \hen}

\subsection{Star formation}

We now discuss the relationships between the 3 main star-forming
regions in \hen.

B is by far the least active region: it has no MIR emission and only 
weak \Hal\ emission.  This lack of MIR emission can be due either to a 
lack of hot stars (such as B stars, that do not produce ionizing 
emission yet are still quite efficient at heating the dust) or of 
dust.  Since Conti \& Vacca (1994) have seen stars clusters in region~B with 
a total UV luminosity equal to 1/4 that of region~A, we believe the 
second hypothesis to be the correct one.  The lack of dust is also 
consistent with the absence of CO emission in region~B (Baas et al., 
\cite{baas}). What is then the relationship between region~A and B?

CKV have suggested that region~A may have been the progenitor of 
region~B, with star formation self-propagating from the former to the 
latter.  We believe that this scenario is not consistent with our MIR 
observations, since we would then expect B to be currently forming 
stars.  In that scheme, the weakness of the \Hal\ emission from B 
would have to be explained by a large dust extinction, but then we 
should have observed a large MIR flux, which is not the case.  
Furthermore, the location of region~B outside of the CO cloud does not 
argue in favor of a large extinction on that line of sight.  Region~A 
could still have triggered region~B if the burst duration in B was 
much shorter than that in A. But this is unlikely given that the 
oxygen abundance is 4 times larger in B than in A (Vacca \& Conti 
1992).  To produce 4 times as much heavy elements than A in a shorter 
time, the burst in region~B would have had to be much more intense 
than in region~A, an unlikely possibility given that the gas density 
around region~B seems to be much lower than toward region~A. Thus it 
is more natural to think that region~B, whose $(B-V)$ color is 
consistent with a population of mainly B stars, is older than 
region~A. That B is older than A is also consistent with the burst 
ages derived from the \Hbet\ equivalent widths measured by Vacca \& 
Conti (1992) and using the model of Viallefond \& Thuan (1983).  The 
age difference of only $\sim$4~Myr, close to the shortest lifetime of 
supernovae progenitors, is probably too small to make plausible the 
hypothesis of a burst in A triggered by a burst in B. The location of 
the CO cloud between the two regions also does not support the 
hypothesis of shock waves propagating from B to A.

\begin{figure*}[ht]
\hbox{
\centerline{
\epsfxsize=6cm
\epsfbox{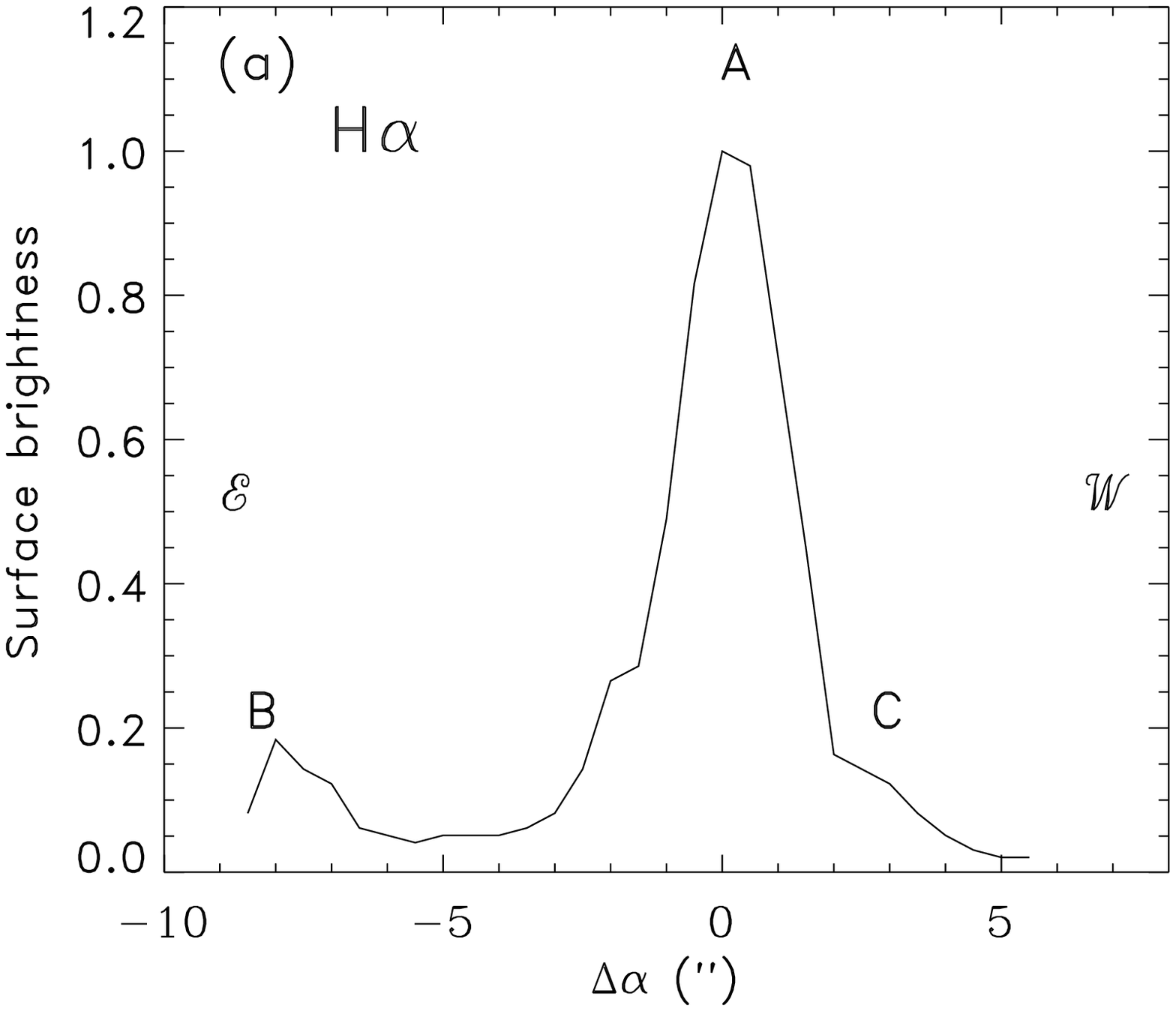}
\epsfxsize=6cm
\epsfbox{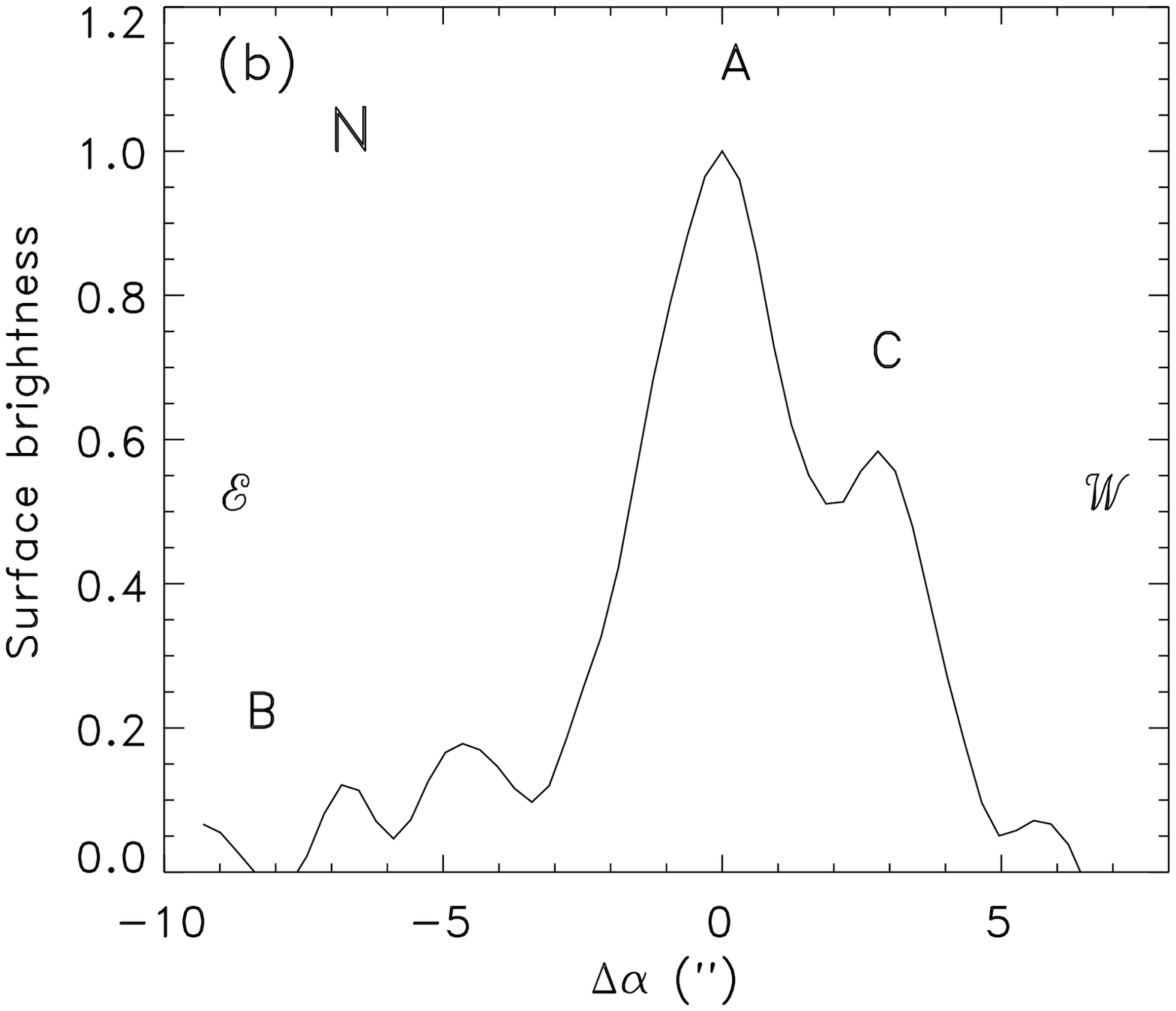}
\epsfxsize=6cm
\epsfbox{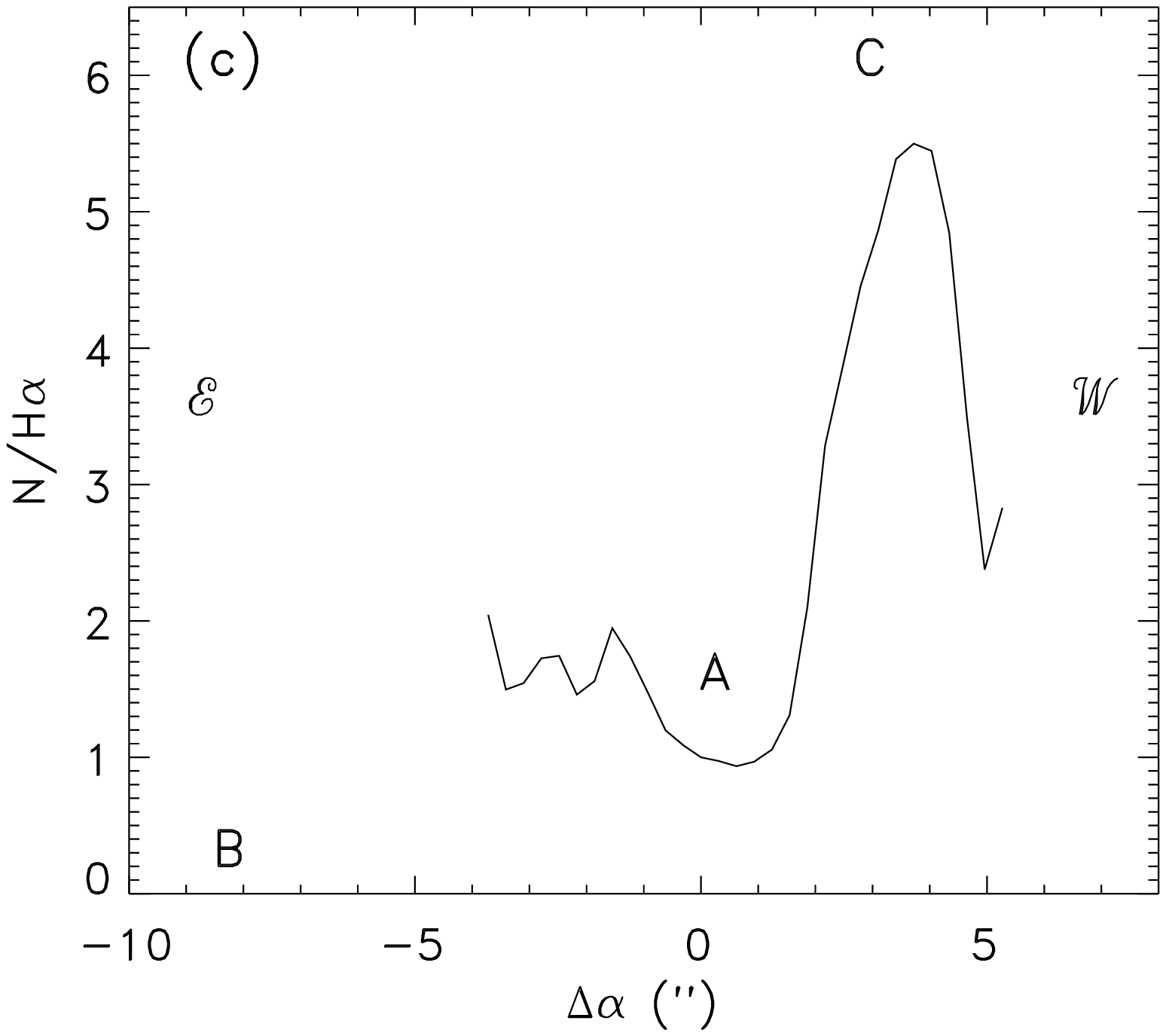}
}
}

\caption[]{Surface brightness profiles in (a) \Hal\ (from ST), (b) N, 
and (c) variation of the ratio N/\Hal\ along an EW cut as defined in 
ST. The profiles are normalized to 1 at the location of region~A. The 
N band data have been convolved to the resolution of the \Hal\ data.}

\label{prof}
\end{figure*}

Because of its relative faintness in $B$ and $V$ images as compared to 
region~A, region~C has not often been discussed in the litterature.  
Yet it shows up clearly in \Hal\ and even more distinctly in [O{\sc 
iii}] (CKV).  We have shown in figure \ref{prof}a the variation of the 
\Hal\ surface brightness along an EW cut through regions A, B and C as 
given by ST. The peak surface brightness of C is $\sim$0.15 times that 
of A, and is similar to that of B. Figure~\ref{prof}b shows the N 
surface brightness along the same cut and convolved to the same 
resolution, as derived from our images.  Region~C is now much brighter 
than B, which is not detected, with a peak surface brightness 
$\sim$0.6 times that of A. Figure~\ref{prof}c shows the resulting 
variation of the N/\Hal\ flux ratio, where region~C has a N/\Hal\ 
ratio $\sim$5.5 times greater than region~A.

This change in the N/\Hal\ ratio can \aprio\ be interpreted as 
indicating a higher extinction (by $\sim$2.5~mag in $V$) in region~C 
than in A, decreasing the \Hal\ emission relative to the N emission.  
However, we reject that hypothesis because (1) the \Hal/\Hbet\ surface 
brightness profile given by ST suggests that the extinction is lower 
in C than in A and (2) the CO column density is much lower toward 
region~C than toward region~A.

A more likely interpretation is that C is smaller and younger than 
region~A. Because it is smaller, it probably contains a smaller number 
of ionizing stars than region~A, and hence has a smaller \Hal\ flux.  
But because it is younger, its heating spectrum contains relatively 
more UV photons than region~A. Because of the shape of the absorption 
curve, these UV photons are more efficiently processed into 
mid-infrared photons by the dust, hence the larger N/\Hal\ ratio.  
That region~C has a higher fraction of ionizing stars than region~A 
is consistent with the higher \Hal\ equivalent width 
$W_{\lambda}(H\alpha)$ of C, $W_{\lambda}(H\alpha)$ varying according 
to the ratios 1.75:1.25:1 for regions C, A and B.

Region~A is the dominant star-forming region of \hen.  More than 2/3 
of the total MIR radiation comes from it.  Vacca \& Conti (1992) 
estimate from the \Hal\ flux that it contains $\sim$5000~O stars and 
300~WR stars.  These estimates are about a factor of 6 smaller than 
those derived from the UV~2200~\AA\ luminosity by Conti \& Vacca 
(1994).  Given the very different light distributions, point-like in 
\Hal\ and extended in the UV, even after taking into account the 
different resolutions, this discrepancy implies that the \Hal\ 
emission comes only from a fraction of the UV knots.  For example, if 
several of the UV knots are older than $\sim$10~Myr, they would still 
produce appreciable $L_{\lambda}(2200 \AA)$, but little Lyman 
continuum and hence \Hal\ emission (Conti \& Vacca 1994).

\subsection{Dust and the interstellar medium}
\label{sec:dust}

We now use the MIR fluxes and colors to discuss the energy budget and
nature of the dust in \hen.

To see whether the starburst in \hen\ can power the whole infrared 
emission of the galaxy, we shall use the Rosette nebula as a template.  
Cox et al.  (\cite{cox90}) derive a bolometric stellar luminosity of 
2.23~10$^{6}$~L$_{\odot}$.  This estimate should be unaffected by 
extinction as it is based on the spectral types of the cluster stars.  
A FIR luminosity of 8.51~10$^{5}$~L$_{\odot}$ is derived from IRAS 
maps giving a ratio of bolometric stellar luminosity to far infrared 
luminosity of $\simeq$2.6\,.  If we adopt the same ratio for \hen, 
we can reproduce the FIR luminosity of the galaxy: O5 stars in the 
\Hal\ emitting UV clusters in region~A plus a 20 percent extra 
contribution from region~C (see fig.\ref{prof}a) would account for a 
FIR luminosity of $\sim1.6\times10^{9}$~L$_{\odot}$.  As for the rest 
of the UV clusters (those that do not emit ionizing radiation), they 
would account for a FIR luminosity of 
0.5-1.2$\times10^{9}$~L$_{\odot}$.  This results in a total of 
2.1-2.8$\times10^{9}$~L$_{\odot}$ while the FIR luminosity of the 
galaxy is 2.6$\times10^{9}$~L$_{\odot}$.  The starburst regions 
present in \hen\ thus seem able to power the infrared emission of the 
galaxy.

Since the global FIR luminosity of \hen\ can be understood as powered 
mainly by the stars in the starburst regions, it is interesting to 
check whether these stars, O and early B stars mainly, can also 
account for the IR spectrum.  To this end, we have used the 
3-component dust models of D\'esert et al.  (1990, the three 
components are PAHs, very small grains, and large grains) where the 
infrared spectrum is computed for various intensities of the heating 
field, this field being either that of an O5 star, that of a B3 star 
or the solar neighborhood one.  We find that none of these models is 
able to reproduce the whole spectrum.  Since the FIR emission is 
consistent with reprocessed OB stellar light, we first tried to make 
the O5 and B3 models fit the long-wavelength part of the IR spectrum.  
The result is that too much emission is predicted in the MIR range (by 
a factor 3 for the best fitting model).  This could indicate that, 
relative to its large grain population responsible for the FIR 
emission, \hen\ lacks small grains and PAH carriers because of its low 
metallicity (see the discussion in Sauvage et al., \cite{sauvageMC}).

To reproduce the short wavelength part of the spectrum with the 
models, we find that we systematically predict less FIR emission than 
what is actually observed in \hen.  This may indicate an additional 
heating source, older stars for example, for the FIR emitting dust.  
However this hypothesis is somewhat in contradiction with the 
observation that the FIR emission can be deduced from the bolometric 
luminosity of the O and B stars in the starburst, assuming a 
conversion factor similar to that in the Rosetta nebula.  It could be 
reconciled with that observation if (1) the true conversion ratio from 
O and B stars luminosity to FIR luminosity is lower in \hen\ than in 
the Rosetta nebula, presumably because of a lower global dust 
abundance in \hen, and (2) the additional heating source by older stars 
is just the right amount as to make the observed ratio of bolometric 
OB stars luminosity to FIR luminosity equal to that observed in the 
Rosetta nebula. 

Another possible explanation for the failure of the model to account 
for the observed spectral distribution is the reprocessing of the 
infrared emission inside the nearby molecular cloud, similar to the 
situation in M~17 (Gatley et al., \cite{gatley}): the copious amounts 
of MIR radiation generated close to the UV clusters penetrate the 
molecular clouds and, depending on the MIR optical depth, a 
significant fraction can be absorbed by dust and re-radiated in the 
FIR. To be efficient, this process requires a large amount of 
material, but the peak \Av\ of 30 measured by Kobulnicky et al.  
(\cite{kobul}) in the CO cloud indicates that this may be 
the case.  As mentionned above, if the model of D\'esert et 
al.  (\cite{dbp}) is to produce the FIR luminosity (60 and 100 \mic), the 
MIR (12 and 25 \mic) luminosity would be overestimated by a factor of 
$\simeq$3.  If we now require that the model only predicts $1/x$ of 
the observed FIR luminosity ($L_{\rm FIR}^{pred}=L_{\rm 
FIR}^{obs}/x$), where x varies between 1 and 3, the 
predicted MIR flux will be overestimated by $\simeq 3/x$ ($L_{\rm 
MIR}^{pred}=3/x \times L_{\rm MIR}^{obs}$).  To obtain the observed MIR 
luminosity, a fraction of $(1-x/3)$ of the predicted luminosity has to be 
reprocessed in-situ to the FIR. In that case, the emerging FIR 
luminosity would be $L_{\rm FIR}^{pred} + (1-x/3) \times L_{\rm 
MIR}^{pred}$.  Given that in the model $L_{\rm FIR}^{pred}/L_{\rm 
MIR}^{pred} = 1.44$, one can solve for x so that the emerging FIR 
luminosity is equal to the observed one. We obtain $x = 1.4$\,.  
The optical depth required to reprocess 54\% of the MIR luminosity
is $\tau_{\rm MIR} = 0.8$.  This value of $\tau$ has to be taken as an 
effective optical depth through the MIR band and is therefore less 
than the optical depth in the silicates band.  Adopting \Av/$\tau_{\rm 
Si} = 19 \pm 1$ (Rieke \& Lebovsky, \cite{rieke}), we need \Av 
$\ga$~15.  It thus appears that internal reprocessing of the dust 
luminosity is the most likely explanation for the failure of the 
models of D\'esert et al (\cite{dbp}) to reproduce the observed 
infrared spectrum.  A more definitive statement would require detailed 
radiative transfer calculations.

\begin{figure}[ht]
\epsfxsize=9cm
\epsfbox{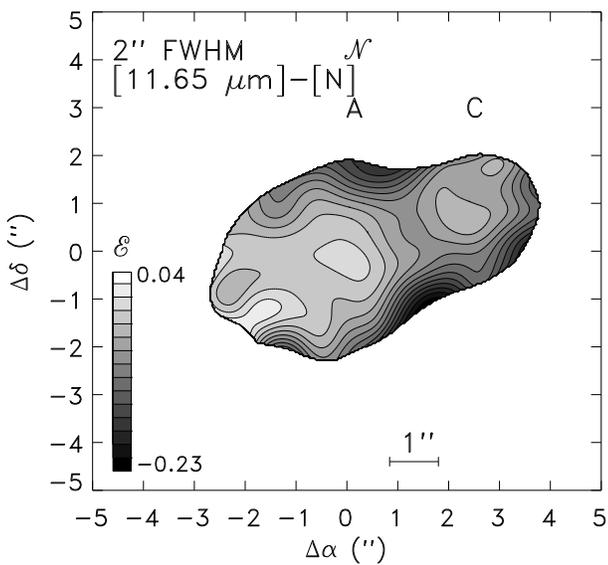}

\caption[]{[11.65~\mic]-[N] color map of \hen.  Only the pixels with 
S/N$\ge$6 in both bands were used to produce the map.  The color is 
minimum on the edges of the map ($\sim-0.23$) and maximum ($\sim0.0$) 
at the location of the MIR emission peak.  The contour interval is 
0.02.}

\label{color}
\end{figure}

We now discuss how the [11.65~\mic]-[N]\footnote{Here [X]$ = \log 
f_{\nu}({\rm X})$.} color map of \hen\ shown in figure~\ref{color} can 
constrain available MIR dust emission models.  To construct the color 
map, we have used only the parts in the 11.65~\mic\ and N images with 
a $S/N\ge6$.  Each image was smoothed with the PSF of its counterpart, 
thus the final resolution is a convolution of the ESO and CFH PSFs 
(approximately 2$''$ FWHM).  It can be seen in figure~\ref{color} that 
the [11.65~\mic]-[N] color varies from $-0.23$ at the edges to 0.0 at 
the location of the peak fluxes.  The higher values ($\sim0.03$) 
toward the eastern edge of the map are probably an artefact of the 
interpolation algorithm used to superpose the individual maps when 
computing the color map.  The mean [11.65~\mic]-[N] color of \hen\ is 
$-0.06$.

To interpret the color map, we have built a simple synthetic model of 
the MIR spectrum made up by the sum of: 1) a continuum parameterized 
by $n$ as defined by $f_{\nu}\propto\nu^{-n}$.  This continuum has 
no physical basis and is simply a convenient parametrization to 
characterize the variations of the spectrum. 2) A silicate 
absorption band using the absorption efficiency of Draine \& Lee 
(1984) and parameterized by the dust column density.  3) PAH emission 
bands, parameterized by the ratio of the 11.3~\mic\ band flux to the N 
flux density, and 4) the NeII line, parameterized by the ratio of its 
flux to the N flux density.  The mean values for parameters 2, 3, and 
4 were obtained from Phillips et al.  (1984) and Roche et al.  (1991), 
while $n$ was varied to obtain the best fit to the observed mean 
color.  We get $n=-1$.

A priori, the task of deriving any useful information from the color 
map seems like a hopeless one, since we have a 4-parameter model for 
only two observational contraints, the mean color and its spatial 
variation.  Fortunately, the last three parameters either induce color 
variations which are not in agreement with the observations or have 
little influence on the color.  The color variation cannot be 
explained by a variation in the dust column density, since to 
reproduce an increase of the [11.65~\mic]-[N] color towards the edges, 
the dust column density would have to be higher at the outskirts of 
the galaxy than at the center, which is nonsensical.  This is due to 
the spectral shape of the silicate absorption coefficient, which 
decreases less steeply longward of 9.7\,\mic\ than shortward of it.  
As a result, increasing the dust column density decreases 
preferentially the 11.65~\mic\ flux density.

As for the PAH emission bands present throughout the N bandpass, since 
in our simple model their ratios to the 11.3~\mic\ band are fixed, 
they do not affect the [11.65~\mic]-[N] color very much.  Multiplying 
the flux of all the PAH bands by a factor of 100 results in a change 
of only $-0.05$ in the color.  Relative band intensity variations could 
also be responsible for the color gradients.  In the extreme case 
where only the 11.3~\mic\ band is present, one still has to multiply 
its flux by 100 to produce a color variation of $\sim0.2$.  Such a 
multiplicative factor is highly unlikely, given the relative constancy 
of the 11.3/N flux ratio from galaxy to galaxy (Roche et al., 1991).  
Furthermore, it would be impossible to have colors smaller than the 
mean [11.65~\mic]-[N] value ($\sim$--0.06), given that the PAH flux is 
such a small fraction of the total flux in both the 11.65~\mic\ and N 
bands (the flux ratio 11.3/N is 3.2$\times10^{-2}$ in \hen, Roche 
et al., 1991).  The same reasoning applies to the NeII flux, the ratio 
of the latter to the N band flux being only $\sim5.3\times10^{-2}$.

It thus appears that the shape of the ``continuum'' as parameterized by 
its slope $n$ is the main factor controlling the color variation.  
Indeed, the full span of the colors seen in figure~\ref{color} can be 
reproduced by varying $n$ from -3 at the outskirts of the map to 0 at 
the location of region~A's peak.  How are we to interpret this slope 
variation? 

If we interpret it in terms of a single dust component made of 
thermalized grains, we run into a problem: the colors at the location 
of the peak MIR fluxes which are highest would be produced by dust at 
the coolest temperature, while the dust at the edges would be the 
hottest, a result which is \underline{not} physical.  The 
``continuum'' slope variation could be explained by the existence of 
an increased emission at long wavelengths in the central regions of 
the map.  This emission could come either from Very Small Grains 
(VSGs, D\'esert et al., 1990) or from warm coal grains (Papoular et 
al., 1993), i.e.  a dust component not related to the PAH band 
carriers.  This extra component would have its peak emission around 
\vtm\ but would still noticeably affect emission toward 10~\mic.  As 
the intensity of the radiation field increases, the temperature of 
this extra dust component increases and its peak emission shifts to 
shorter wavelengths.  Given that emission shortward of 10~\mic\ is 
dominated by PAH bands, the net result of the higher radiation field 
would be a relative increase of the total dust emission on the 
long-wavelength side of the N band, mimicking the change in slope 
noted before.  Such an interpretation has already been brought 
forward to explain the MIR colors of M~82 (see e.g.  Telesco et al., 
\cite{telesco90}, their figure 13) and those of NGC~253 (Telesco et 
al., \cite{telesco93}).  It can also be clearly seen in the 
spectra and images of star-forming regions obtained by ISO (see e.g.  
Vigroux et al., \cite{vigroux96}).

Alternatively, these ``continuum'' slope variations could be 
explained by PAH-related variations. Indeed it is clear that even if 
the bands themselves contribute a relatively small fraction of the 
total MIR flux, some of them seem to be associated with much broader and 
intense features. This is the case of the 7.7 and 8.6~\mic\ bands 
which are located on a very strong ``bump'' that is probably 
physically connected to them and that is providing most of the flux at 
these wavelengths. Thus the  observed slope variation could be due to a 
variation of this bump's relative contribution to the N-band flux 
density, 
its contribution being dominant in the outskirts of our map. This 
is plausible as in these outer regions, the radiation field is 
probably  too low for emission other than that related to PAH 
carriers to be significant.

Finally we note that both mechanisms described above can actually be 
at work in \hen: in the central regions of the map, the UV flux is so 
intense that emission from very small grains appears in the 
11.65~\mic\ band and contributes to a flattening of the MIR spectrum.  
Away from these UV sources, this emission disappears from the 
11.65~\mic\ band while impulsive heating becomes the dominant 
mechanism and results in a spectrum dominated by the strong PAH blend 
around 8~\mic.  Furthermore, the mean index for the color of \hen\ 
implies that this blend is a dominant feature of the spectrum, 
a property that could be related to the high mean UV flux 
density found in \hen.  These conclusions can actually be checked by 
mapping this feature with ISO in Blue Compact Dwarf galaxies.

Since the above simple modelling of the spectral energy distribution 
in the MIR gives us a slope of the continuum emission in the 
8-13\,\mic\ range, we can examine how our and the IRAS measurement 
compare.  The \dzm\ flux density of 1.1\,Jy corresponds to a band-integrated 
flux of 1.5\,10$^{-13}$\,W\,m$^{-2}$.  This is just the value we 
obtain by using our measured N~band flux density and convolving the 
spectral ernergy distribution $\nu^{+1}$ with the IRAS bandpass. The 
good agreement implies that most of the MIR emission comes from 
the star-forming regions and that there is not a significant extended 
MIR diffuse halo, which would have been detected with the much larger 
beam of IRAS.

\section{Conclusions}

In this paper, we have presented the first high-resolution MIR images 
of the Blue Compact Dwarf galaxy \hen, as well as a new measurement of 
its \Hi\ content.  Table~1 lists the general properties of \hen.  Our 
conclusions can be summarized as follows:

\begin{enumerate}

\item The galaxy is detected and resolved into sub-components in the 
N-band and 11.65~\mic\ filters.  The main component, region~A, at the 
center of the galaxy, is elongated in the SE-NW direction and has the 
same position angle in both MIR images.  The same shape and 
orientation are also seen in the optical.  The arc of UV clusters seen 
by Conti \& Vacca (1994), and the \Hal\ emitting region detected by 
Sugai \& Taniguchi (1992) are contained entirely within the MIR 
emission of region~A. The latter emits $\sim2/3$ of the MIR flux of 
the galaxy.  The second bright optical region of \hen, region~B, at 
8$\farcs$5 E of region~A, is not detected.  A third region, C, 
appearing in the optical as a small \Hii\ region 2$\farcs$7 W of region 
A, is clearly detected.  Both regions~A and C are at the edge of a 
very dense CO cloud. 

\item The present star formation activity is very different from one 
region to the other.  Region~A is the most active, as is evident from 
its high \Hal, UV and MIR fluxes.  It is younger than region~B: though 
the latter contains UV emitting stars, ionizing stars have died and 
most of the interstellar medium has been exhausted, as evidenced by 
the lack of CO emission.  We attribute the lack of MIR emission to a 
lack of dust rather than to a lack of heating flux.  Region~C, 
although smaller and less luminous in \Hal\ than region~A contains a 
higher proportion of ionizing to non-ionizing stars, as shown by its 
N/\Hal\ ratio, which is 5.5~times higher than that of region A. We 
show that this difference cannot be attributed to a higher extinction, 
and probably reflects the younger age of region~C.

\item We found that the large-beam \dzm\ IRAS measurement is similar 
to our narrow-beam measurements, implying that there is no diffuse MIR 
emission.  The UV-emitting clusters provide enough energy to heat all 
the dust emitting in the FIR. We show that the models of D\'esert et 
al.  (\cite{dbp}) are however unable to fully account for both the 
observed MIR and FIR spectra.  The model predicts too much MIR 
emission relative to the FIR emission.  One possible way for 
reconciling the models' predictions with the observations is to 
invoke reprocessing of the MIR emission into FIR emission by dust 
close to the UV clusters, inside the very large molecular complex to 
the west of the starburst, where a very high extinction has been 
found.

\item We found comparable estimates for the dust mass using either the 
60 and \ctm\ IRAS flux densities or the silicate absorption depth.  In both 
cases, the dust mass is $\sim8~10^{4}$~M$_{\odot}$.  The corresponding 
dust-to-gas mass ratio is $\sim16$~times smaller than in the solar 
neighborhood but is typical of other BCDs.

\item The MIR color variations imply either that, in addition to PAH 
band carriers, there is an extra-component of very small grains with 
peak emission at $\sim$ 25\,\mic\ contributing in the hottest regions 
of the galaxy, or that the PAH ``blend'' of the 7.7 and 8.6~\mic\ band 
strongly increases relative to the rest of the MIR emission toward the 
outskirts of the galaxy. A combination of these two mechanisms is 
actually quite likely.  Furthermore, the mean color of \hen\ is 
characteristic of a spectrum where this PAH ``blend'' at short 
wavelengths (7.7 and 8.6\,\mic) is dominant relative to the 
11.3\,\mic\ feature.  This is probably due to the high UV flux 
density found in \hen.

\item Our new \Hi\ measurement shows that \hen\ has a gas mass 
fraction on the low side of the range for BCDs.

\end{enumerate}

\acknowledgements{CAMIRAS and TIMMI were developed with the invaluable 
help of Ren\'e Jouan and Pierre Masse at Saclay.  We thank them for 
their diligence and efficiency, especially at the high altitude of the 
CFHT. Eric Pantin's skills at deconvolution were greatly appreciated.  
We are grateful to Renaud Papoular for helpful discussions on silicate 
and coal dust.  T.X.T. thanks Laurent Vigroux for his hospitality at 
the Centre d'Etudes de Saclay.  M.S. thanks Suzanne Madden for 
providing the [CII] measurement of \hen\ in advance of publication, as 
well as for discussions on its significance.  We also acknowledge 
significant inputs from our referee, Dr.  C.M. Telesco.  This research 
has made use of the SIMBAD database, operated by the Centre de 
Donn\'ees Stellaires in Strasbourg, and of the NASA/IPAC Extragalactic 
Database (NED) which is operated by the Jet Propulsion Laboratory, 
Caltech, under contract with the National Aeronautics and Space 
Administration.}

\end{document}